# Machine learning methods in quantum computing theory

D.V. Fastovets*[ab], Yu.I. Bogdanov[abc], B.I. Bantysh[ab], V.F. Lukichev[a]

[a]Valiev Institute of Physics and Techonology of Russian Academy of Sciences, Russia, Moscow; [b]National Research University of Electronic Technology (MIET), Russia, Moscow; [c]National Research Nuclear University (MEPhI), Russia, Moscow

## ABSTRACT

Classical machine learning theory and theory of quantum computations are among of the most rapidly developing scientific areas in our days. In recent years, researchers investigated if quantum computing can help to improve classical machine learning algorithms. The quantum machine learning includes hybrid methods that involve both classical and quantum algorithms. Quantum approaches can be used to analyze quantum states instead of classical data. On other side, quantum algorithms can exponentially improve classical data science algorithm. Here, we show basic ideas of quantum machine learning. We present several new methods that combine classical machine learning algorithms and quantum computing methods. We demonstrate multiclass tree tensor network algorithm, and its approbation on IBM quantum processor. Also, we introduce neural networks approach to quantum tomography problem. Our tomography method allows us to predict quantum state excluding noise influence. Such classical-quantum approach can be applied in various experiments to reveal latent dependence between input data and output measurement results.

**Keywords:** quantum computing, qubits, quantum algorithms, machine learning, artificial intelligence

## 1. INTRODUCTION

Machine learning algorithms are tasked with extracting main information and making prediction about new data samples. In contrast to other mathematical techniques, these algorithms construct and update their predictive model based on known data (training dataset). Machine learning methods can be applied to different tasks: spam filter, image processing, wide societal impact, image recognition, signal processing and etc. [1-3].

In recent years, there have been a lot of advances in the quantum information processing showing that particular quantum algorithms can provide a speedup over their classical analogues [4,5]. It has been shown that application of quantum approaches to the field of classical machine learning may produce similar results. Such combination of quantum computing power and machine learning ideas would be a great boost to the quantum information science field and may evaluate new practical solutions for current machine learning problems [6-9].

Quantum computing has efficient advantage in multi-dimensional systems and multi-variable statistical analysis [10,11]. Classical systems with a large number degrees of freedom are difficult to modeling because of the curse of dimensionality [12]. But quantum parallelism effect allows us to avoid this problem. Therefore, quantum computational resources are very useful to solve different problems with extra-large number of dimensions (machine learning, for example).

There are many methods associated with data processing and machine learning procedure. Quantum approaches can be applied to basic methods (k-nearest neighbor algorithm, k-means algorithm, principal component analysis and etc.) as well as to sophisticated methods (variational autoencoders, associative adversarial networks and etc.). Here, we demonstrate applicability of quantum information techniques to both classes of algorithms.

The structure of the present paper is as follows. The basic quantum algorithms helpful in machine learning and the computing disadvantages of famous quantum machine learning (QML) algorithms are presented in Section 2. It is shown that quantum algorithms have more efficiency than classical only for data with very large dimension.

*fast93@mail.ru

The quantum classification circuit, which called tree tensor networks (TTN), presented in Section 3. Here, the TTN generalization for multiclass classification is presented. It is based on softmax function idea and allows us to predict label for data with arbitrary number of classes. Real approbation on IBM online quantum processor was performed and corresponding results are shown.

Neurotomography (quantum tomography based on classical artificial neural networks) is presented in Section 4. Here we use neural network model to predict output quantum state excluding noise in quantum system. Such approach can be applied in real experimental systems to reconstruct correct quantum state with high probability.

The key results of the paper are stated briefly in Section 5.

## 2. REAL CHARACTERISTICS OF QUANTUM ALGORITHMS

### 2.1 SWAP-test

At first, we describe basic quantum algorithms which will be used in machine learning problems. The *SWAP-test* routine is a simple quantum algorithm expressing scalar product of two input states $|a\rangle$ and $|b\rangle$ (Fig. 1). This algorithm has been firstly used in [13]. It is necessary to get a lot of measurements to reach correct value of scalar product between two states.

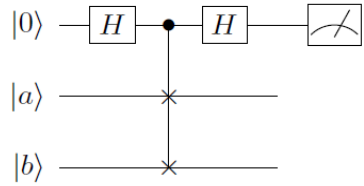

Figure 1. Quantum circuit of SWAP-test algorithm.

The probability of measuring control qubit being in state $|0\rangle$ is given by:

$$P(|0\rangle) = \frac{1}{2} + \frac{1}{2} F(|a\rangle, |b\rangle), \qquad (1)$$

where $F(|a\rangle, |b\rangle) = |\langle a|b\rangle|^2$ - fidelity between two input states. The probability $P(|0\rangle) = 0.5$ means that the states $|a\rangle$ and $|b\rangle$ are orthogonal. And the probability $P(|0\rangle) = 1$ indicates that the states are identical. The routine should be repeated several times ($\geq 10^4$) to obtain a good estimation of fidelity.

Suppose we had two quantum states: $|\psi\rangle = \frac{1}{\sqrt{2}}(|0, a\rangle + |1, b\rangle)$ and $|\varphi\rangle = \frac{1}{\sqrt{Z}}(|a||0\rangle + |b||1\rangle)$, where $Z = |a|^2 + |b|^2$ - normalization factor. Fidelity between these quantum states associated with scalar product of vectors $|a\rangle$ and $|b\rangle$. Using of SWAP-test for these quantum states we can obtain the scalar product value:

$$|a-b|^2 = 2Z\, F(|\psi\rangle, |\varphi\rangle) = 2Z(2P(|0\rangle) - 1). \qquad (2)$$

SWAP-test can be used to calculate Euclidian distance between vectors in multidimensional space. It is very useful for various classical algorithms.

### 2.2 Classical data encoding

The classical *d*-dimensional vector $a$ is encoded in quantum state as [14]:

$$|a\rangle = \frac{1}{|a|} \sum_{i=1}^{d} a_i |i\rangle, \qquad (3)$$

where $|i\rangle$ - quantum state corresponding to binary representation of $i$. Therefore, $d$-dimensional vector can be translated into $\log_2 d$ qubits. The computational complexity of quantum states preparation to calculate scalar product (2) is $O(\log d)$ [15].

### 2.3 Quantum minimization algorithm (QMA)

Quantum algorithm for solving discrete functions minimization problems was introduced in [16]. It is based on Grover's search and can be represented in terms of quantum iterations circuits (Fig. 2).

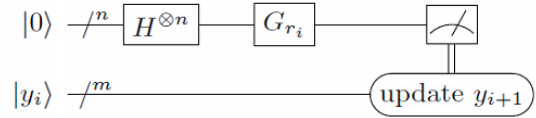

Figure 2. Quantum circuit of quantum minimization algorithm.

Firstly, we randomly select point $x_1$ and set threshold $y_1 = f(x_1)$ using minimized function $f$. Next, we apply Hadamard (see Fig. 2) gates to zero register. Then, we apply Grover iteration $r_1$ times and measure first $n$ qubits. Measurements results allow us to correct threshold value. Therefore, we start second iteration: set threshold $y_2$, apply Hadamard gates and Grover's iteration $r_2$ times, measure first qubits and update threshold again (all $r_i$ are choosed randomly). The convergence of this iteration procedure is achieved in order $\sqrt{2^n}$ iterations. Such algorithm allows us to find global optimum as it is always searching through all possible inputs states.

### 2.4 K-nearest neighbors (KNN) algorithm

KNN is a very popular and simple classification algorithm. Given a training dataset $T$ of feature vectors and corresponding class labels. The idea of algorithm is to classify new data (test dataset $D$) by choosing the class label for each new input vector that appears most often among its $k$ nearest neighbors. Such idea is based on the simple rule: close feature vectors have the similar class labels. It is true for many applications. The used modeling datasets $T$ and $D$ are presented in Fig. 3.

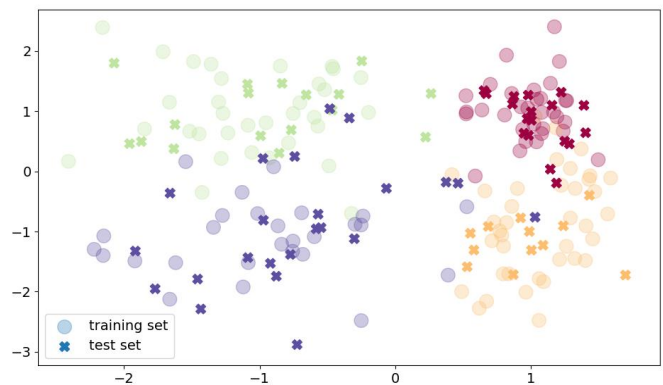

Figure 3. Training and test datasets (4 classes) used for modeling quantum KNN algorithm.

Quantum version of KNN is based on using SWAP-test (to calculate all scalar products between feature vectors encoded in quantum states (3)) and QMA (to get $k$ closest neighbors for each test vector). Therefore, quantum algorithm efficiency depends on number of SWAP-test iterations ($n_{ST}$) and number of QMA iterations ($n_{QMA}$). Our modeling results (for datasets presented on Fig. 3) are shown on Fig. 4. Classification accuracy is demonstrated using confusion matrix for different values $n_{ST}$ and $n_{QMA}$.

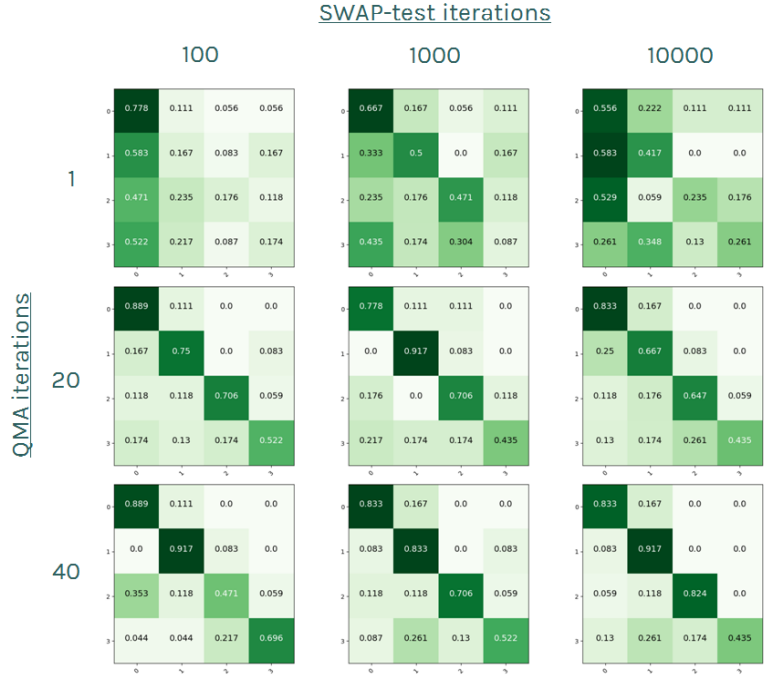

Figure 4. Confusion matrices for different values of SWAP-test iterations and QMA iterations.

Let $n$ - training set size, $m$ - test set size, $k$ - number of clusters and $d$ - space dimension. Then, classical KNN computational complexity is $O(nm(k+d))$. Quantum approaches can improve this value. The quantum KNN complexity is $O(nm(k+\log d))$. Here we can see the computational advantage. But if we take into account the constants $n_{ST}$ and $n_{QMA}$ we will see the dependences presented on Fig. 5.

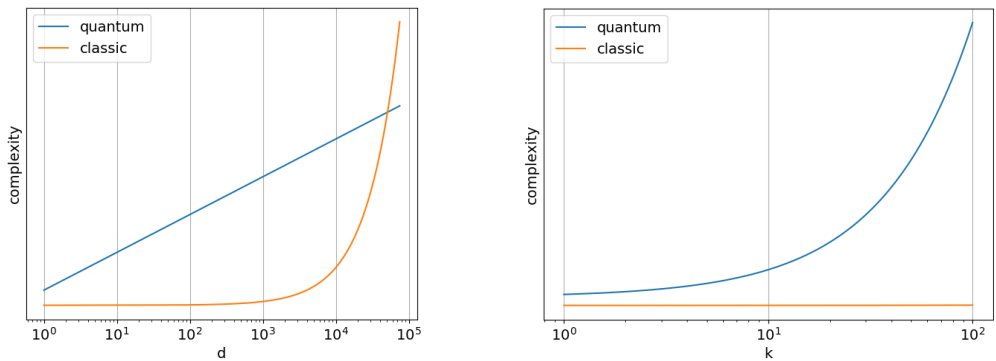

Figure 5. KNN algorithm computational complexities. Left - $k < d$, right - $k > d$.

As we can see, quantum approach leads to efficient computation only for $d > 5 \cdot 10^4$ (Fig. 5, left picture). Therefore, quantum KNN algorithm can be applied to the task with large space dimension.

**2.5 K-means algorithm**

Classical *k*-means algorithm is one of the basic unsupervised machine learning algorithms. This algorithm classify data to *k* clusters base on unlabeled set of vectors. The vectors from set relabeled to the nearest centroid label in each iteration and then a new centroid is calculated by averaging all current cluster vectors. The quantum version of *k*-means is based on using SWAP-test (to calculate distances) and QMA (to calculate closest centroid). For example, we used the dataset presented on Fig. 6.

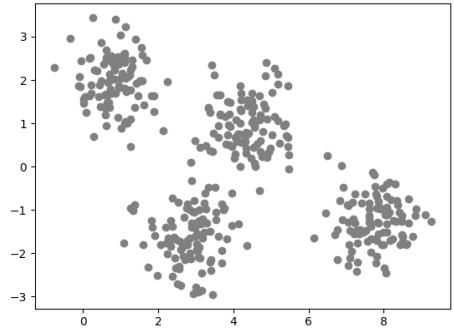

Figure 6. Unlabeled dataset with 4 classes.

Our clusterization results (using quantum version of *k*-means algorithm) are shown on Fig. 7.

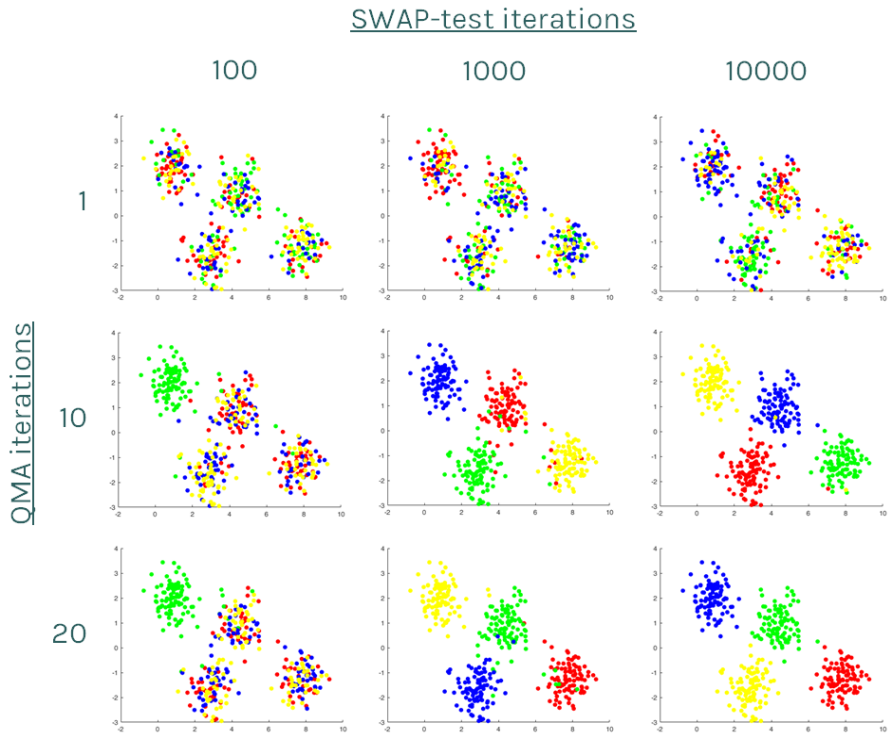

Figure 7. Clusterization results for different values of SWAP-test iterations and QMA iterations.

Using quantum algorithms provides an exponential speed-up in comparison to classical *k*-means algorithm. The classical version of algorithm executed on a classical computer with complexity $O(knd)$, where $k$ – number of clusters, $n$ - dataset size, $d$ - space dimension. This evaluation corresponds to the fact that for each vector we need to calculate distances to all $k$ clusters. The time complexity of the quantum algorithm is $O(kn\log d)$. This complexity give speed-up for large dimensions of vectors. To plot complexity dependences (Fig. 8) we use constants $n_{ST}$ and $n_{QMA}$ to get real complexity value.

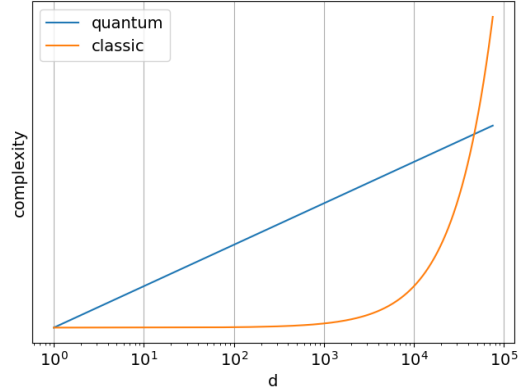

Figure 8. K-means algorithm computational complexity.

As we can see, for a small space dimension it could be advisable to use classical algorithm.

## 3. TREE TENSOR NETWORKS

### 3.1 Data encoding

Let us first consider the case of classical data for binary classification problem. It is a set $D = \{(x_i, y_i)\}_{i=1}^{n}$ where $x_i \in \mathbb{R}^d$ are $d$-dimensional input vectors, and $y_i \in \{0,1\}$ are the corresponding class labels. Classifying classical data with quantum algorithm requires encoding input vectors to quantum states. The most efficient approach in terms of space is (3). However, such method is helpful for modeling only because there is no known efficient method that can prepare quantum state in the form (3). A simpler method is to encode each element of a classical vector in the single-qubit amplitude. This type of encoding requires $d$ qubits to encode $d$-dimensional data vector. However, the state preparation procedure requires single-qubits rotations only. Before encoding, we need to rescale the data vectors to $\left[0, \frac{\pi}{2}\right]$. Then, we encode each data vector element in a qubit using the following formula

$$|\psi_i\rangle = \cos(x_i)|0\rangle + \sin(x_i)|1\rangle. \tag{4}$$

The final quantum state is $|\psi\rangle = \otimes_{i=1}^{n} |\psi_i\rangle$.

### 3.2 Binary classification algorithm

The tree tensor networks (TTN) is an architecture of quantum circuits inspired by binary trees [17]. The TTN circuit begins by applying unitary transforms to each qubits pairs. Then, we discard one of the qubit from each pairs. In the following circuit layer we again apply two-qubit gates to the remaining qubits pairs. This process is repeated until only one qubit remains. Such architecture for 8 input qubits presented in Fig. 9.

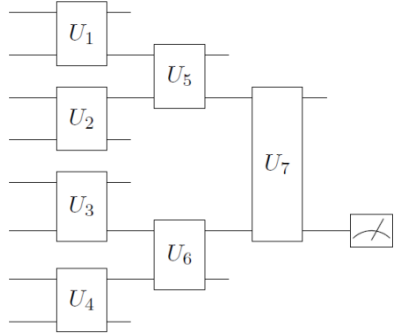

Figure 9. TTN quantum circuit architecture.

Measuring remains qubit gives us $|0\rangle$ or $|1\rangle$. Several shots (for fixed input vector $x_i$) of this circuit allow us to obtain probability $P(|0\rangle)$. If $P(|0\rangle) < 0.5$ we assign label $\tilde{y}_i = 0$ to source vector $x_i$, $\tilde{y}_i = 1$ otherwise. It is a binary classification.

To speed-up learning process (it is provides by classical computer), we need to use unitary parameterization. One of the simplest way to do that is use CNOT gate and two one-qubits rotations, around Y axis, for example (because rotation around Y axis doesn't commute with CNOT and consist from real values in matrix representation). The efficiency of TTN was demonstrated on Iris flower dataset. It is a four-dimensional dataset introduced by Ronald Fisher. The data consists of 50 samples from each of three species of iris (iris setosa, iris verginica and iris versicolor). Four features were measured from each sample: the length and the width of the sepals and petals, in centimeters. The quantum circuit for four-dimensional input vectors (using unitary parameterization) is presented in Fig. 10.

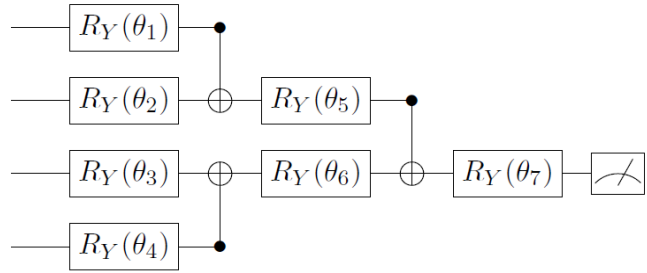

Figure 10. Parameterized TTN quantum circuit for Fisher's iris dataset.

All parameters $\theta = \{\theta_1, \theta_2, \theta_3, \theta_4, \theta_5, \theta_6, \theta_7\}$ are optimized to correctly classify training dataset (usually represented by 70% of the original dataset). Such result obtains by loss function minimization

$$L(\theta) = \sum_{i=1}^{r} (\tilde{y}_i - y_i)^2, \quad (5)$$

where $r$ is a training set size. Loss function usually optimized by stochastic gradient descent. It is a stochastic because at each iteration the gradient is estimated on a small batch rather on the full training set. Because we use binary classification algorithm, we will explore pairwise classification: setosa vs verginica, setosa vs versicolor and verginica vs versicolor. The training processes are presented on Fig. 11.

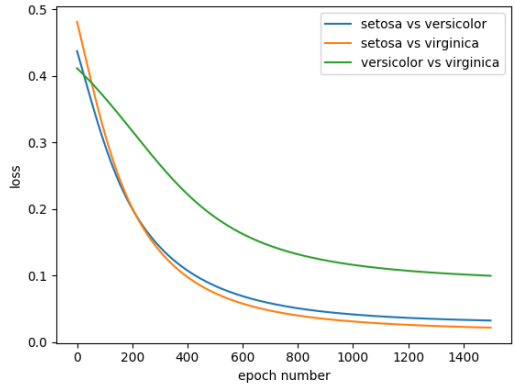

Figure 11. Training process: loss function optimization.

Using other $n-r$ vectors (test set) allows us to obtain accuracy of models. Accuracy is a basic metric for evaluating classification models. Therefore, accuracy is the fraction of predictions our model got right. In our case, accuracy between setosa and versicolor classes is $A=1$, between setosa and virginica classes is $A=1$ too, and between versicolor and virginica classes is $A=0.94$. The last result is due to corresponding classes are inseparability in feature space.

### 3.3 IBM implementation

The circuit (Fig. 10) with obtained parameters $\theta$ can be implemented on IBM online quantum processor (ibmqx4). Encoding procedure (4) is represented by single-qubit rotation U3 for each qubit.

Table 1. TTN IBM implementation.

|  | Number of shots | | | | | | | |
|---|---|---|---|---|---|---|---|---|
|  | 1 | 3 | 5 | 9 | 21 | 55 | 201 | 1001 |
| Setosa vs Versicolor | 0.818 | 0.879 | 0.879 | 0.939 | 0.970 | 1 | 1 | 1 |
| Setosa vs Virginica | 0.666 | 0.818 | 1 | 1 | 1 | 1 | 1 | 1 |
| Versicolor vs Virginica | 0.606 | 0.606 | 0.727 | 0.789 | 0.848 | 0.879 | 0.909 | 0.909 |

### 3.4 Multiclass classification algorithm

Here, we introduce the generalization of TTN binary classificator. We will use analogue of softmax function to predict labels. Lets condiser the following quantum circuit based on TTN idea (Fig. 12).

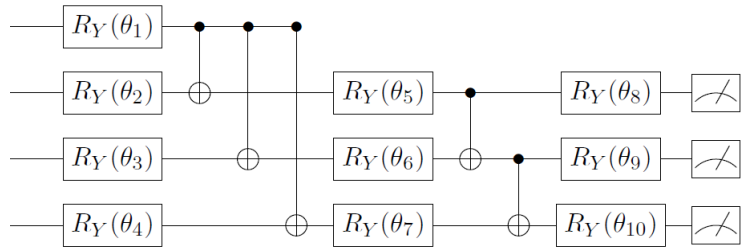

Figure 12. Generalized TTN circuit for 3-class classification.

We can obtain three probabilities after measurements (Fig. 12) $P_1(|0\rangle), P_2(|0\rangle), P_3(|0\rangle)$. Then, we train our model to predict class $\tilde{y}=\arg\max(P_1(|0\rangle), P_2(|0\rangle), P_3(|0\rangle))$. Therefore, qubit index with maximal zero-state probability provide a class number. Here, we can construct confusion matrix to present classification results:

Table 2. Confusion matrix for 3-class classification.

|  | Setosa | Versicolor | Virginica |
| --- | --- | --- | --- |
| Setosa | 1 | 0 | 0 |
| Versicolor | 0 | 1 | 0 |
| Virginica | 0 | 0.09 | 0.91 |

## 4. NEUROTOMOGRAPHY

Quantum tomography is the process of reconstructing the quantum state by measurement results [18-20]. We introduce a quantum state tomography method based on neural networks approach. Our fully connected neural net was trained to convert measurements results to the parameters of pure one-qubit state (polar and azimuth angles on the Bloch sphere).

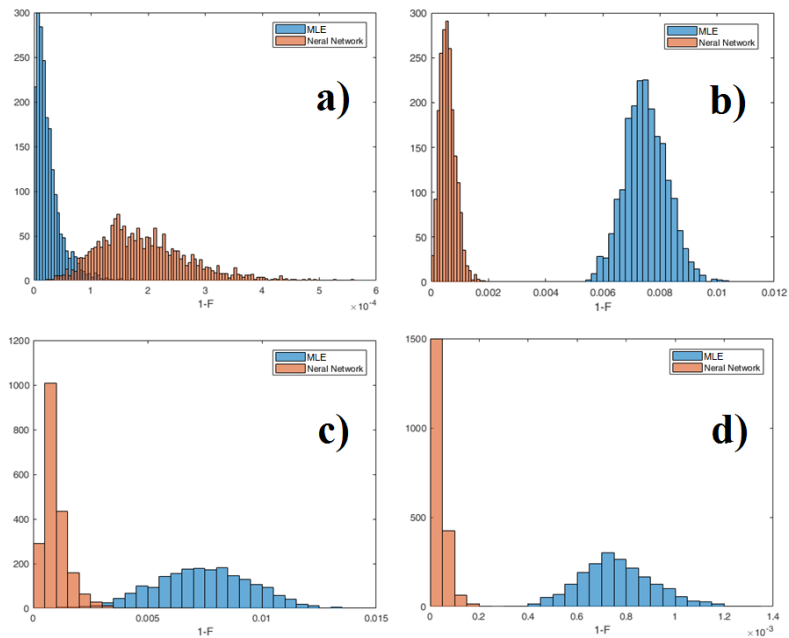

Figure 13. Neurotomography and MLE tomography comparison. Number of experiments: 2000. Sample size: $10^5$.

The basic tomography methods, developed in our laboratory [21,22] (based on maximum likelihood estimation (MLE)) work better then neural network approach (Fig. 13a). But it is true for ideal quantum systems. Neurotomography allows us to reconstruct quantum state amplitudes in noisy quantum systems also (without knowing the noise model).

As we can see on Fig. 13, neural network trained on noisy dataset ($10^4$ quantum states uniformly distributed over Bloch sphere) gives the better result than classical tomography approach. The presented results weakly dependent on noise type: systematic unitary error (Fig. 13b), random unitary error (Fig. 13c) or amplitude-phase relaxation (Fig. 13d). Neural network reveals the hidden relationship between quantum states and noisy measurements result.

# 5. CONCLUSIONS

Our results are presented the approaches to quantum machine learning methods. We introduce tree tensor network (TTN) quantum circuit that can be used as binary classifier. We demonstrate efficiency of these methods on basic classical dataset (Fisher's Iris). Also we demonstrate TTN approach on the IBM quantum processor.

Here, we demonstrate quantum tomography with using neural networks approach. Such classical-quantum approach can be applied in various experiments to reveal latent dependence between input data and output measurement results.

## ACKNOWLEDGMENTS

This work was supported by Russian Foundation of Basic Research (project 18-37-00204).

## REFERENCES

[1] Bishop, C.M., [Pattern recognition and machine learning], Springer, 738 (2006).
[2] Goodfellow, I., Bengio, Y. and Courville, A., [Deep learning], The MIT Press, 800 (2016).
[3] Mitchell, T.M., [Machine learning], McGraw-Hill, 432 (1997).
[4] Mosca, M., "Quantum algorithms", arXiv:0808.0369, 1-71 (2008).
[5] Coles, P.J., Eidenbenz, S., Pakin, S. et al. "Quantum Algorithm Implementations for Beginners", arXiv:1804.03719, 1-76 (2018).
[6] Schuld, M., Sinayskiy, I. and Petruccione, F., "An introduction to quantum machine learning", Contemp. Phys., 56(2), 172-185 (2015).
[7] Cai, X.-D., Wu, D., Su, Z.-E., Chen, M.-C., Wang, X.-L., Li, L., Liu, N.-L., Lu, C.-Y. and Pan, J.-W. "Entanglement-Based Machine Learning on a Quantum Computer", Phys. Rev. Lett., 114(2), 110504 (2015).
[8] Adcock, J.C., Allen, E., Day, M. et al. "Advances in quantum machine learning", arXiv:1512.02900, 1-38 (2015).
[9] Kopczyk, D., "Quantum machine learning for data scientists", arXiv:1804.10068, 1-46 (2018).
[10] Kendall, M.G. and Stuart, A. [The advanced theory of statistics, volume 3: Design and analysis and time-series] Charles Griffin & Company, 567, (1968).
[11] Hart, G.W., [Multidimensional analysis. Algebras and systems for science and engineering], Springer, 236, (1995)
[12] Sammut, C. and Webb, G.I. [Encyclopedia of machine learning and data mining], Springer US, 1335 (2017).
[13] Aumeur, E., Brassard, G. and Gambs, S., "Machine learning in a quantum world", Proc. Conference of the Canadian Society for Computational Studies of Intelligence, 431-442 (2006).
[14] Vartiainen, J.J., Bergholm, V. and Salomaa, M.M. "Transformation of quantum states using uniformly controlled rotations", Quant. Inf. Comp., 5, 467 (2005).
[15] Lloyd, S., Mohseni, M. and Rebentrost, P., "Quantum algorithms for supervised and unsupervised machine learning", arXiv:1307.0411, 1–11 (2013).
[16] Christoph, D. and Hoyer, P. "A quantum algorithm for finding the minimum", arXiv:quant-ph/9607014, 1-2, (1996).
[17] Shi, Y., Duan, L. and Vidal, G., "Classical simulation of quantum many-body systems with a tree tensor network", Phys. Rev. A., 74, 022320 (2006).
[18] Rehacek, J., Englert, B.-G. and Kaszlikowski, D., "Minimal qubit tomography", Phys. Rev. A., 70, 052321 (2004).
[19] Burgh, M.D., "Choice of Measurement Sets in Qubit Tomography", Phys. Rev. A., 78, 052122 (2007).
[20] D'Ariano, G.M., Paris, M.G.A. and Sacchi, M.F., "Quantum Tomography", Adv. Imaging Electron Phys., 128, 205–308 (2003).
[21] Bogdanov, Yu.I., Brida, G., Genovese, M., Kulik, S.P., Moreva, E.V. and Shurupov, A.P. "Statistical estimation of the quality of quantum-tomography protocols", Phys. Rev. A - At. Mol. Opt. Phys, 84(4), 1–19 (2011).
[22] Bogdanov, Yu.I., Bantysh, B.I., Bogdanova, N.A., Kvasnyy, A.B. and Lukichev, V.F., "Quantum states tomography with noisy measurements channels", Proc. SPIE, 10224 (2016).